# Citations: Indicators of Quality? The Impact Fallacy

Loet Leydesdorff *[a], Lutz Bornmann,[b] Jordan Comins,[c] and Staša Milojević[d]

**Abstract**

We argue that citation is a composed indicator: short-term citations can be considered as currency at the research front, whereas long-term citations can contribute to the codification of knowledge claims into concept symbols. Knowledge claims at the research front are more likely to be transitory and are therefore problematic as indicators of quality. Citation impact studies focus on short-term citation, and therefore tend to measure not epistemic quality, but involvement in current discourses in which contributions are positioned by referencing. We explore this argument using three case studies: (1) citations of the journal *Soziale Welt* as an example of a venue that tends not to publish papers at a research front, unlike, for example, *JACS*; (2) Robert K. Merton as a concept symbol across theories of citation; and (3) the Multi-RPYS ("Multi-Referenced Publication Year Spectroscopy") of the journals *Scientometrics, Gene*, and *Soziale Welt*. We show empirically that the measurement of "quality" in terms of citations can further be qualified: short-term citation currency at the research front can be distinguished from longer-term processes of incorporation and codification of knowledge claims into bodies of knowledge. The recently introduced Multi-RPYS can be used to distinguish between short-term and long-term impacts.

**Keywords**: citation, symbol, historiography, RPYS, obliteration by incorporation

[a] University of Amsterdam, Amsterdam School of Communication Research (ASCoR), PO Box 15793, 1001 NG Amsterdam, The Netherlands; loet@leydesdorff.net ;
[b] Division for Science and Innovation Studies, Administrative Headquarters of the Max Planck Society, Hofgartenstr. 8, 80539 Munich, Germany; email: bornmann@gv.mpg.de
[c] Center for Applied Information Science, Virginia Tech Applied Research Corporation, Arlington, VA, United States; email: jcomins@gmail.com
[d] School of Informatics and Computing, Indiana University, Bloomington 47405-1901, United States; smilojev@indiana.edu

* corresponding author: Loet Leydesdorff, loet@leydesdorff.net

## 1. Introduction

When asked about whether citations can be considered as an indicator of "quality," scientometricians are inclined to withdraw to the position that citations measure "impact." But how does "impact" differ from "quality"? Whereas Cole & Cole (1973, p. 35), for example, argued that "the data available indicate that straight citation counts are highly correlated with virtually every refined measure of quality," Martin & Irvine (1983) claimed that quality is indicated only in cases where several indicators converge (e.g., numbers of publications, citations, etc.), thus introducing the notion of partial indicators. In their view "the indicators based on citations are seen as reflecting the impact, rather than the quality or importance, of the research work" (Martin & Irvine, 1983, p. 61). Moed *et al.* (1985), on the other hand, framed the discussion of the relationship between "impact" and "quality" in the context of enabling science-policy decisions so as to distinguish research groups in terms of their visibility and their longer-term "durability"; that is, their potential to make sustained contributions to a field of science in terms of short-term citation impacts during a longer period of time.

With the increase of usage of quantitative indicators for evaluation of individuals, groups, universities, and nations, revisiting the relations between "quality" and "impact" is both timely and important. The operationalization of quality in terms of impact leads first to the question of the definition of "impact." "Impact" is a physical metaphor used by Garfield & Sher (1963, p. 200) when introducing the "journal impact factor" (JIF). Unlike size-dependent indicators of impact based on the total number of citations (Gross & Gross, 1927), the impact *factor* normalizes for the size effects of journals by using a (lagged) two-year moving average.

Scientometricians distinguish between size-dependent and size-independent indicators. Analytically, one would expect "quality"—as against "quantity"—to be size-independent, whereas "impact" is size dependent, since two collisions have more impact than a single one. Leydesdorff & Bornmann (2012) argued for an indicator based on integrating impact instead of averaging it in terms of ratios of citations per publication. Bensman & Wilder (1968) found that faculty judgements about the quality of journals in chemistry correlate empirically more with total citation rates than with (size-independent) impact factors. Bensman (2007, p. 118) added that Garfield had modeled the journal impact factor on the basis of an early version of the SCI in the 1960s (Garfield, 1972, p. 476; Martyn & Gilchrist, 1968) in which bio-medical journals with rapid yearly citation turnover would have been dominant.

In the meantime, scientometricians have become thoroughly aware that (*i*) publication and citation practices differ among disciplines; and (*ii*) one should not use the average of sometimes extremely skewed distributions (Seglen, 1992), but should instead use non-parametric statistics (e.g., percentiles). In their recent guidelines for evaluation practices, Hicks *et al.* (2015, p. 430), for example, conclude that "(n)ormalized indicators are required, and the most robust normalization method is based on percentiles … in the citation distribution of its field." However, the definition of percentiles presumes reference sets or, in other words, the demarcation of "fields" of science. The top-10% can be very different from one reference set to another.

In evaluative bibliometrics, a "best practice" has been developed to delineate reference sets in terms of three criteria: cited publications should be (*i*) from the same year (so that they have had equal opportunity to gather citations); (*ii*) of the same document type (articles, reviews, or letters, so that documents of the same depth and structure can be compared); and (*iii*) from the same field of science, each of which has its own distinct citation patterns. The first two criteria are provided by the bibliographic databases,[1] but the delineation of fields of science has remained a hitherto unresolved problem (Leydesdorff & Bornmann, 2016; van Eck *et al*., 2013). Although one can undoubtedly assume an epistemic structure of disciplines and specialties operating in the sciences, the texture of referencing can be considered as both woofs and warps: the woofs may refer, for example, to disciplinary backgrounds and the warps to current relevance (Quine, 1960, p. 374). Decomposition of this texture using one clustering algorithm or another may be detrimental to the evaluation of units at the margins or between fields (e.g., Rafols *et al*., 2012), and the effects are also sensitive to the granularity of the decomposition (Waltman & van Eck, 2012; Zitt *et al.*, 2005).

Perhaps, these can be considered as technical issues. More fundamentally, the question of normalization refers to differences in citation behavior among fields of science (Margolis, 1967). Wouters (1999) argued that the use of citations in evaluations is first based on the transformation of the citation distribution from "citing" to "cited": the citation indexes collect cited references—which are provided by citing authors/texts—into aggregated citations. Such a transformation of one distribution ("citing") into another ("cited") is not neutral: papers may be cited in fields other than those they are citing from. Whereas documents are cited, citation behavior is an attribute of authors. The "cited-ness" distribution can be used out of context (e.g., for rankings) and thus apart from the reasons for "citing" (Bornmann & Daniel, 2008).

Does this abstraction legitimate us to compare apples (e.g., excellently elaborated texts) with oranges (e.g., breakthrough ideas)? Normalization brings the citing practices back into the design because one tries to find reference sets of papers cited for similar reasons or in comparable sets. However, the reasons for citation may be very different even within a single text (Amsterdamska & Leydesdorff, 1989; Chubin & Moitra, 1975; Moravcsik & Murugesan, 1975). The assumption that journals, for example, contain documents which can be compared in terms of citation behavior abstracts from the reasons and the content of citation by using the behavior of authors as the explanatory variable.

Citation counts may seem convenient for the evaluation because they allow us to make an inference *prima facie* from "quality" in the textual to the socio-cognitive dimensions of authors and ideas, or vice versa (Leydesdorff & Amsterdamska, 1990). However, the results of the bibliometric evaluation inform us about the qualities of document sets, and not immediately about authors, institutions (as aggregates of authors), or the quality of knowledge claims. Furthermore, the aggregation rules of texts, authors, or ideas (as units of analysis) are different. For example, a single text can be attributed as credit to all contributing authors or proportionally to the number of authors using so-called "fractional counting;" but can one also fractionate the knowledge claim? A citation may mean something different with reference to textual, social, or epistemic structures.

---

[1] The distinction between review and research articles in the Web-of-Science (WoS) is based on citation statistics: "In the *JCR* system any article containing more than 100 references is coded as a review. Articles in 'review' sections of research or clinical journals are also coded as reviews, as are articles whose titles contain the word 'review' or 'overview.'" at http://thomsonreuters.com/products_services/science/free/essays/impact_factor/ (retrieved Feb. 22, 2016).

At the epistemic level, Small (1978) proposed to consider citations as "concept symbols". Would one be able to use citations for measuring not only the impact of publications and the standing of authors, but also the quality of ideas? Are ideas to be located within specific documents or between and among documents; i.e., in terms of distributions of links such as citations or changes in word distributions? One can then formulate a research agenda for theoretical scientometrics in relation to the history and philosophy of science, but at arm's length from the research agenda of evaluative bibliometrics where the focus is on developing more refined indicators and solving problems of normalization.

In this study, we use empirical findings from a number of case studies to illustrate what we consider to be major issues at the intersection of theoretical and evaluative bibliometrics, and possible ways to move forward. We first focus on sociology as a case with an extremely long turn-over of citation. However, longer-term citation is also important in other disciplines (Ke *et al.*, 2015; van Raan, 2004): short-term citation at the research front can be considered as citation currency, whereas codification of citation into concept symbols is a long-term process. Historical processes tend to be path-dependent and therefore specific. Citation indicators such as the impact factor and SNIP (Moed, 2010), however, focus on citation currency or, in other words, participation at a research front. The extent to which short-term citation can be considered as a predictor of the long-term effects of quality can be expected to vary (Baumgartner & Leydesdorff, 2014; Bornmann & Leydesdorff, 2015).

## 2. German sociology journals

The journal *Soziale Welt*—subtitled "a journal for research and practice in the social sciences"— can be considered as twice disadvantaged in evaluation practices: the discipline (sociology) has a low status in the informal hierarchy among the disciplines[2] and the journal publishes for a German-speaking audience. Special issues, however, are sometimes entirely in English. German sociology has a well-established tradition, and many of the ground-breaking debates in sociology have German origins (e.g., Adorno *et al.*, 1970); but since the Second World War German sociology has mostly been read in English translation (e.g., Schutz, 1967). Merton (1973b) noted that bi-lingual journals serve niche markets in sociology. The special position of German sociology journals enables us to show the pronounced effect on citation patterns and scores of being outside main-stream science.

---

[2] On average, impact factors in sociology are an order of magnitude smaller than in psychology (Leydesdorff, 2008, p. 280).

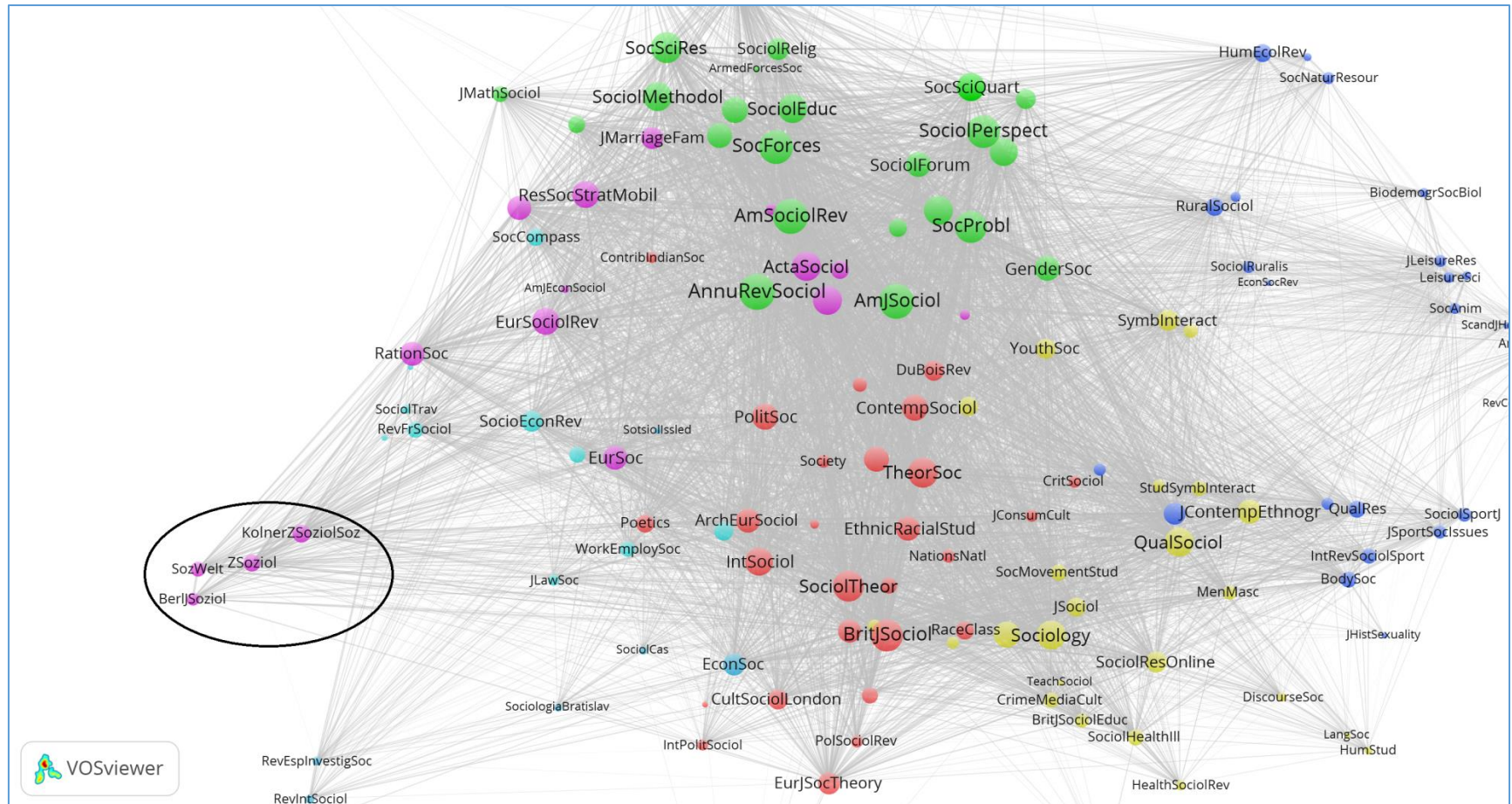

**Figure 1**: Four German sociology journals among 141 journals classified as "sociology" in the Web-of-Science (JCR 2014), mapped in terms of their (cosine-normalized) being cited patterns using VOSviewer for the mapping and the classification.

Figure 1 shows the four German sociology journals included in Thomson-Reuters' Web-of-Science (WoS) when mapped in terms of their "cited" patterns in relations to other sociology journals. The "citing" patterns of these same journals, however, are very different: on the "citing" side the journals are deeply integrated in sociology, which provides the knowledge base for their references. In other words, the identity of these journals is sociological, but their audience is the German-language realm, including journals in political science, education, psychology, etc. Thus, journals can show very different patterns for being cited or citing, and the same asymmetry (cited/citing) holds for document sets other than journals. For example, the *œuvre* of an author or an institutionally delineated set of documents (e.g., a department) cannot be expected to match journal categories. One may be indebted ("citing") to literatures other than those to which one contributes (e.g., Leydesdorff & Probst, 2009).

In the case of these German sociology journals, the border is mainly a language border, but disciplinary distinctions can have the same effect as language borders. The codification of languages ("jargons") in the disciplines and specialties drives the further growth of the sciences because more complexity can be processed in restricted languages (Bernstein, 1971; Coser, 1976; Leydesdorff, 2006). "Translational research" in medicine—the largest granting program of the U.S. National Institute of Health—deliberately strives to counteract these dynamics of differentiation by focusing on translation from the laboratory, with its language of molecular biology, to clinical practice in which one proceeds in terms of clinical trials and protocol development (e.g., Hoekman *et al.*, 2012).

Processes of translation between disciplines or between specialist and interdisciplinary contexts (e.g., *Nature* or *Science*) require careful translation. Interdisciplinary research is not based on a melting pot of discourses, but on the construction of codes of communication in which the more restricted semantics of specialisms can be embedded (Wagner *et al.*, 2009). Asymmetries in the relations among the various discourses lead, among other things, to different citation rates. Which reference set ("field") would one, for example, wish to choose in the case of *Soziale Welt*? The one of its citing identity, or the one in which it is cited?

A second disadvantage of *Soziale Welt* in terms of journal impact factors is the virtual absence of short-term citation that would contribute to its JIF-value. The JIF is based on the past two or five years (JIF-2 and JIF-5, respectively); Elsevier's SNIP index for journals is based on the past three years. In 2014, however, *Soziale Welt* was cited 98 times in WoS, of which 63 (64.3%) were citations of papers published more than ten years ago. However, this slow turn-over is not specific to German sociology journals.

The *American Journal of Sociology* (AJS) and *American Sociological Review* (ASR)—the two leading sociology journals—have cited and citing half-life times of more than ten years. In other words, more than half of the citations of these journals are from issues published more than ten years ago, and more than half of the references in the 2014 volumes were to publications older than ten years. Unlike the German journals, the American journals also have short-term citation which leads to a JIF of 3.54 for AJS, and 4.39 for ASR. These impact factors are based on only 1.9% and 2.7%, respectively, of these journals' total citations in 2014.[7]

**Table 1**: Journals with cited and citing half-life of more than ten years in JCR 2014

|                | N of Journals | Cited Half-Life > 10 | Citing Half-Life > 10 |
|----------------|---------------|----------------------|-----------------------|
| **SCI-Expanded** | 8613          | 1947                 | 2553                  |
|                |               | 18.5%                | 29.6%                 |
| **Social SCI** | 3134          | 779                  | 1473                  |
|                |               | 24.9%                | 47.0%                 |

Table 1 shows that longer-term citation is not marginal in disciplines other than sociology. Almost half of the journals included in the *Social Science Citation Index* (47%) have a citing half-life of more than ten years. In the natural and life sciences, long-term citation is also substantial. The *Journal of the American Chemical Society* (JACS), for example, has a cited half-life time of 8.0 years and a citing half-life time of 6.5 years. However, this journal obtains 14.7% of its citations in the first two years after publication and 37.3% within five years.[8]

## 3. Short-term and long-term citations: citation currency and codification

Let us disaggregate citations at the journal level and examine the long-term and accumulated citation rates of specific—highly-cited—papers in greater detail. Figure 2 shows the ten most highly cited articles in *Soziale Welt.* Nine of these papers were not cited more than four times in any given year. These incidental citations accumulate over time. The single exception to this pattern is Bruno Latour's (1996) contribution to the journal (in English) entitled "On actor-network theory: a few clarifications." Almost ten years after its publication, this paper began to be cited at an increasing rate. From this perspective, all other citations to *Soziale Welt* can be considered as noise. In sum, after a considerable number of years Latour's (1996) paper became a concept symbol (Small, 1978), whereas the other papers remained marginal in terms of their citation rates.

---

[7] The 2012 and 2013 volumes of AJS were cited 63 and 171 times in 2014 out of a total citation count of 12,416. For ASR, the numbers are 101, 259, and 13,181, respectively. For *Soziale Welt,* the percentage of citation to publications in the last two years is 12.2%; (3+9)/98.
[8] The numbers for 2014 are 489,761 total cites; 71,941 as the numerator of IF-2, and 182,760 for IF-5.

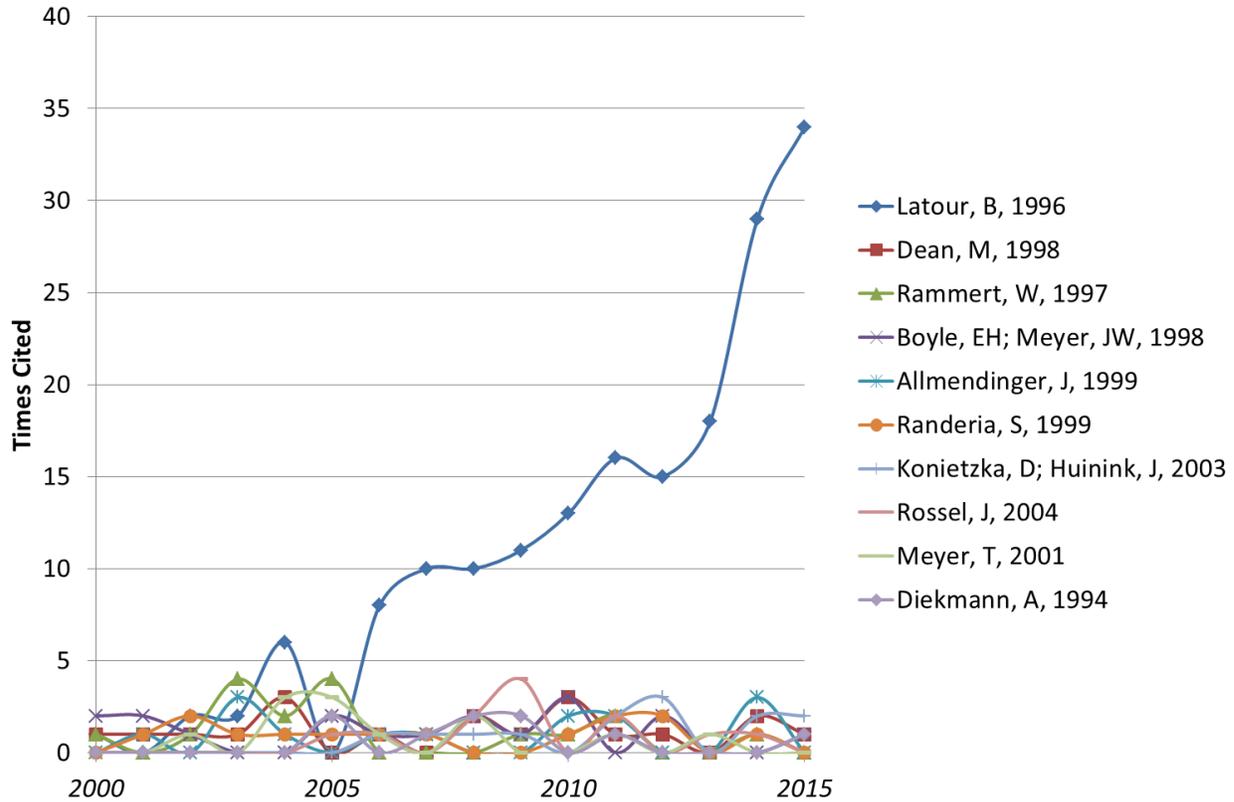

**Figure 2**: Ten most-cited papers in *Soziale Welt* (14 February 2016).

Let us repeat this analysis for the top-10 most highly cited papers in the *American Journal of Sociology,* a core journal of this same field. Figure 3 shows the results: three papers show the deviant behavior which we saw for Latour (1996) in Figure 2. However, these ten papers all have citation rates of more than one thousand times. Whereas the seven at the bottom continue to increase in terms of yearly citation rates over the decades, the top-3 accelerate this pattern with almost twice the rate. Coleman's (1988) study entitled "Social Capital in the Creation of Human-Capital" became a most highly cited paper after almost two decades (since 2008; cf. van Raan, 2004): it went from 61 citations in 2007 to 231 citations in 2008. Note that none of these top-10 papers are decaying in terms of the citation curve, as one would expect given the normal pattern after so many years.

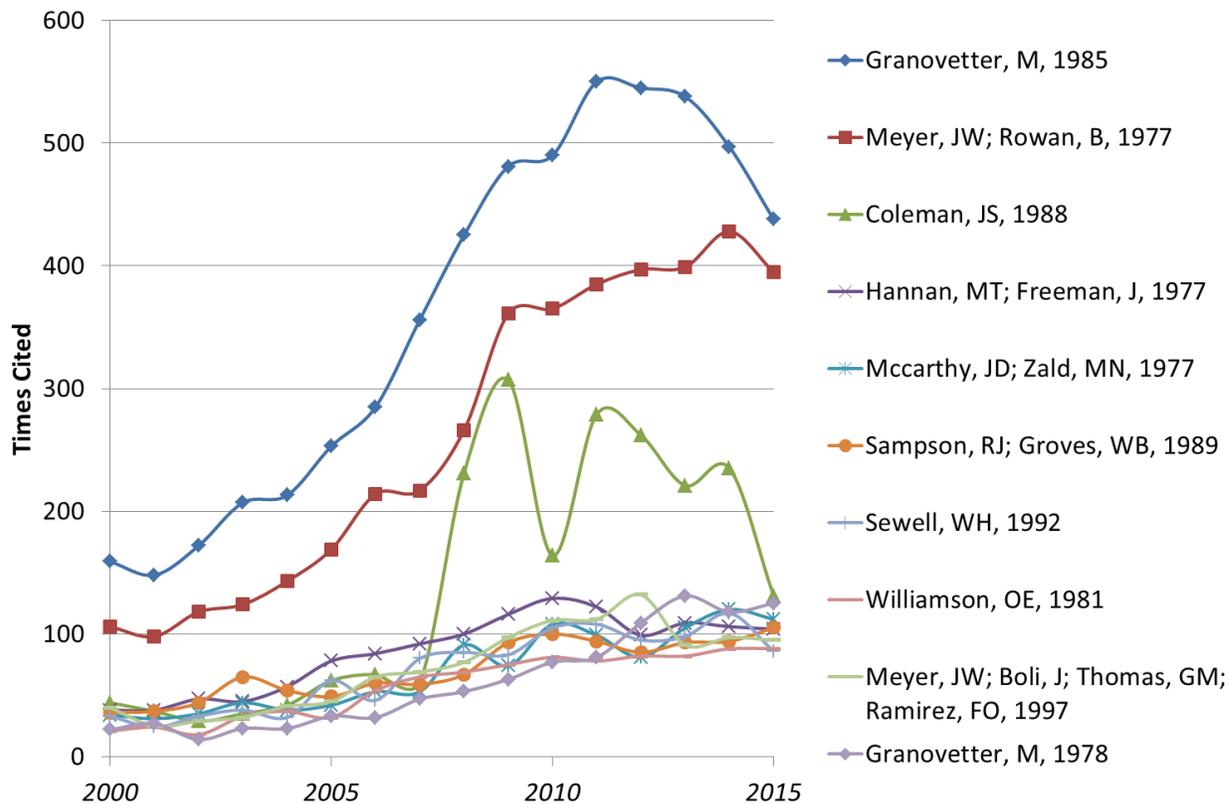

**Figure 3**: Ten most-cited papers in *AJS* (14 February 2016).

Using a similar format, Figure 4 shows the ten most-highly-cited papers in *JACS*. One of these is Kamihara *et al.*'s (2008) paper, entitled "Iron-Based Layered Superconductor La[O1–xFx]FeAs (x = 0.05–0.12) with $T_c = 26$ K". Despite its empirical title, this paper is directly relevant for the theory of superconductivity, and therefore was being cited immediately.[9] The citation curve of this paper shows the standard pattern of a successful contribution: the paper was cited 353 times in the year of its publication, peaked in 2010 with 763 citations, and thereafter the curve decays. Using the IF-type systematic, one can say that it gathered 1,513 of its total of 4,647 citations (32.6%) in the first two years following on its publication, and 3,369 (72.5%) in the first five years. This pattern is typical for a paper at a research front (Price, 1970).

---

[9] The immediacy or Price index is the percentage of papers cited in the year of their publication (Moed, 1989; Price, 1970).

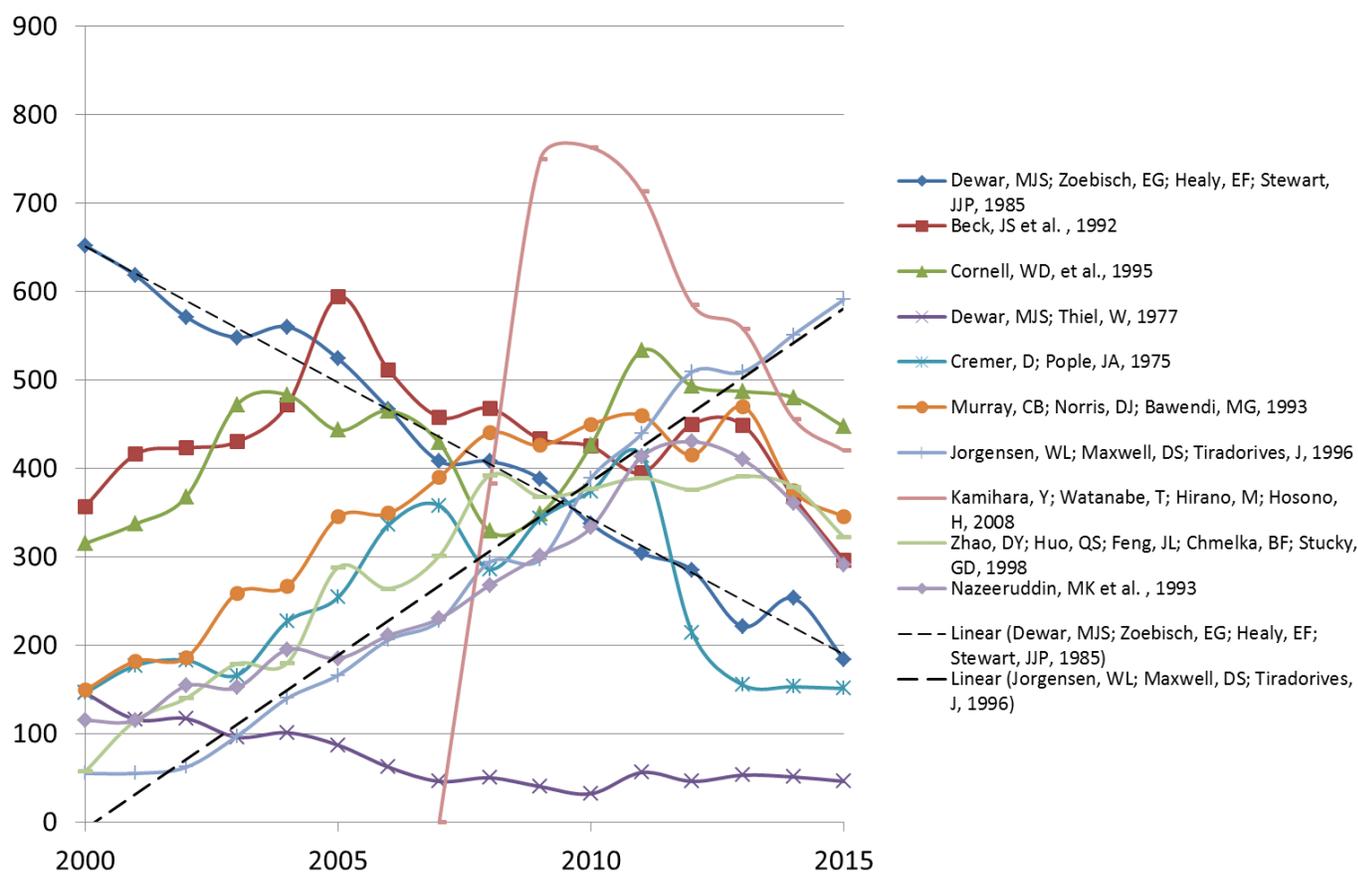

**Figure 4**: Ten most-cited papers in *JACS* (14 February 2016).

The other nine highly-cited papers show different patterns. Some are increasing, while others decrease in terms of yearly citation rates. Jorgensen *et al.* (1996), for example, shows sustained linear growth in citation, whereas Dewar *et al.* (1985) has been decreasing since 1997 when it was cited 826 times (after twelve years!).

Using Group-Based Trajectory Modeling (Nagin, 2005), Baumgartner & Leydesdorff (2014) studied a number of journals, among them *JACS,* in terms of the citation patterns of all papers published in 1996. Figure 5 shows the seven trajectories distinguished among the citation patterns of (2,142) research articles published in *JACS* during 2016, using a 15-year citation window. Although a number of the (statistically significant) groups show typical citation patterns with an early peak and decay thereafter, groups 5 (7.24% of the papers) and 7 (1.31%) were still increasing their citation rates after 15 years. The authors consistently found that the decay phase was not continuous across journals and fields of science (see, for example, group 6 in the middle).[10]

---

[10] A 5[th]- order polynomial was needed for the modeling, indicating that the decay (3[rd] order) is disturbed by other processes of citation behavior.

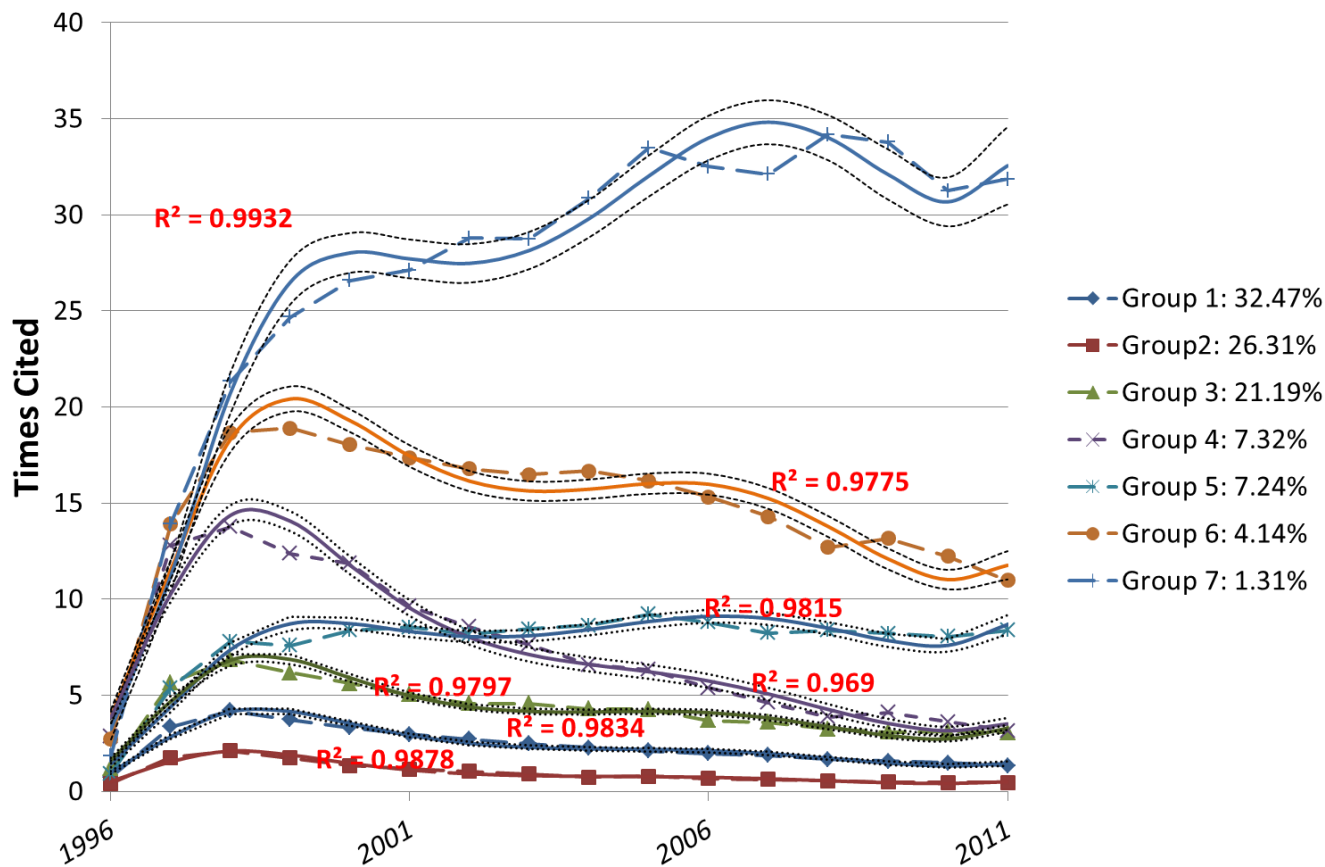

**Figure 5**: Seven trajectories of 2,142 research articles published in *JACS* in 1996, using 5[th]-order polynomials. Source: Baumgartner & Leydesdorff, 2014, p. 802.

Baumgartner & Leydesdorff (2014) proposed to distinguish between "sticky" and "transitory" knowledge claims. Transitory knowledge claims are typical for the research front; the community of researchers informs one another about progress. Sticky knowledge claims need time to grow into a codified citation that can function as a concept symbol (Small, 1978). Evaluation in terms of citation analysis focuses on transitory knowledge claims at the research front. Comins & Leydesdorff (2016 and forthcoming) call this the citation currency of the empirical sciences.

The current discourse at a research front is provided by transitory knowledge claims with variations that contribute to shaping the research agenda at the above-individual level. The attribution of the results of this group effect to individual authors or texts is at risk of the ecological fallacy: part of the success is due to relations among individual contributions, and one cannot infer from quality at the group level to quality at the individual level (Robertson, 1950).[11] The huge delays in citation that we found above in sociology may indicate that generational change is also needed in fields without a research front before a new concept symbol becomes highly cited. Let us focus on one such concept symbol, most central to our field: Robert K. Merton, who among many other things defined the "Matthew effect"—preferential attachment— in science and who is often cited for his "normative" theory of citation (e.g., Haustein *et al.*, 2015; Wyatt *et al.*, 2016).

---

[11] Another example of the ecological fallacy is the use of impact factors of journals as a proxy for the quality of individual papers in these journals (Alberts, 2013).

## 4. "Merton" as a concept symbol across theories of citation

A "citation debate" has raged in the sociology of science between the constructivist and the normative theories of citation (Edge, 1979; Luukonen, 1997; Woolgar, 1991). The normative theory of citation (Kaplan, 1965) is grounded in Merton's (1942) formulation of the CUDOS norms of science: Communalism, Universalism, Disinterestedness, and Organized Skepticism. From a Mertonian perspective, citation analysis can be considered as a methodology for the historical and sociological analysis of the sciences (e.g., Cole and Cole 1973; Elkana *et al.*, 1978; Price, 1965). Citation is then considered as a reward and thus an indicator of the credibility of a knowledge claim.

In a paper entitled "A different viewpoint," Barnes & Dolby (1970) argued for shifting the attention in sociology from the professed (that is, Mertonian) norms of science to citation practices. Gilbert (1977), for example, studied referencing as a technique of rhetorical persuasion, whereas Edge (1979, p. 111) argued that one should "give pre-eminence to the *account from the participant's perspective*, and it is the *citation analysis* which has to be 'corrected'" (italics in the original). The field of science and technology studies (STS) thus became deeply divided between quantitative scientometrics mainly grounded in the Mertonian tradition and qualitative STS dominated by constructivist assumptions (Luukkonen, 1997). During the 1980s, however, the introduction of discourse analysis (Mulkay *et al.*, 1983) and co-word maps (Callon *et al.*, 1983) made it possible to build bridges from time to time (Wyatt *et al.*, 2016; cf. Leydesdorff & van den Besselaar, 1997; van den Besselaar, 2001).

Let us use references to "Merton" as a concept symbol in the citation debate between these theories of citation. Merton can be expected to be cited across the entire set of this literature because proponents as well as opponents use and discuss his ideas. This analysis is based on the full sets of publications in *Scientometrics* (since 1978) and *Social Studies of Science* (since 1971) downloaded from WoS on October 6, 2014 in another context (Wyatt *et al.*, 2016). These are 5,677 publications in total, of which 3,891 were published in *Scientometrics*, and 1,786 in *Social Studies of Science*. [12] These 5,677 records in the document set contain 159,373 references. Among these are 595 references to Merton in 391 documents. In other words, Merton is cited (as a first author)[13] in 6.9% of the documents.

---

[12] The latter figure includes 1,689 published in *Social Studies of Science* (since 1975) and 97 in *Science Studies* (the previous title of the journal between 1971 and 1974).
[13] The cited references in WoS provide only the names and initials of first authors. Citations to Zuckerman with Merton as second author are therefore not included.

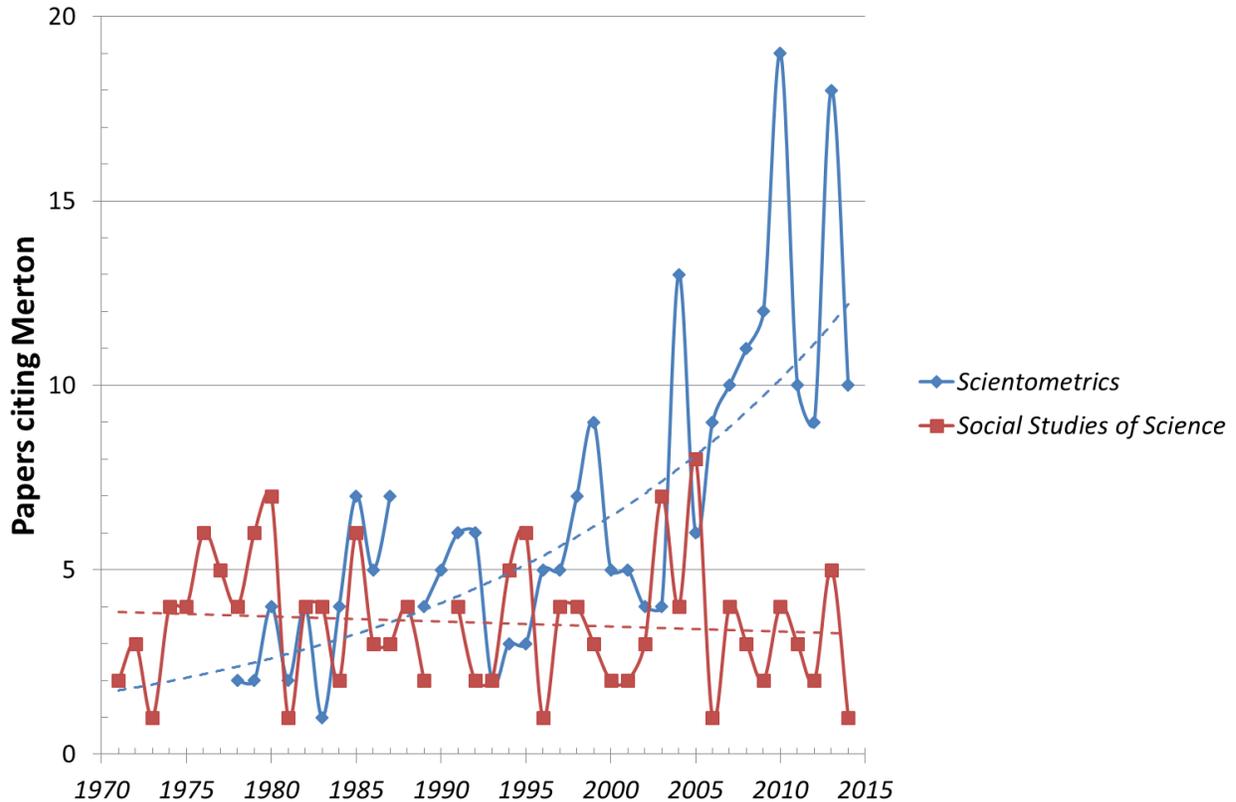

**Figure 6**: Distribution of references to Merton in papers in *Scientometrics* and *Social Studies of Science*, respectively. (*n* of documents = 391).

Figure 6 shows that the number of references to Merton is declining steadily in *Social Studies of Science*, but increases in more recent years in *Scientometrics.* From the perspective of hindsight, Merton's various contributions to institutional sociology can be considered as scientometrics *avant la lettre*. Price's "cumulative advantages" (1976, p. 292), for example, operationalized Merton's (1968) "Matthew effect"—the tendency for citation-rich authors and publications to attract further citations, in part *because* they are heavily cited (Cole & Cole, 1973; Crane, 1969, 1972; Bornmann *et al.*, 2010).[14] The theoretical notions of both Merton and Price thus anticipated the concept of "preferential attachment" in network studies by decades (Barabási & Albert, 1999; Barabási *et al.*, 2002). The mechanism of preferential attachment, for example, enables scientometricians to understand the Matthew effect as a positive feedback at the network level that cannot be attributed to the original author (e.g., Scharnhorst & Garfield, 2011), the journal (Larivière & Gingras, 2010), or the country of origin (Bonitz *et al.*, 1999).

---

[14] The so-called Matthew Effect is based on the following passage from the Gospel: "For unto every one that hath shall be given, and he shall have abundance: but from him that hath not shall be taken away even that which he hath." (Matthew 25:29, King James verrsion).

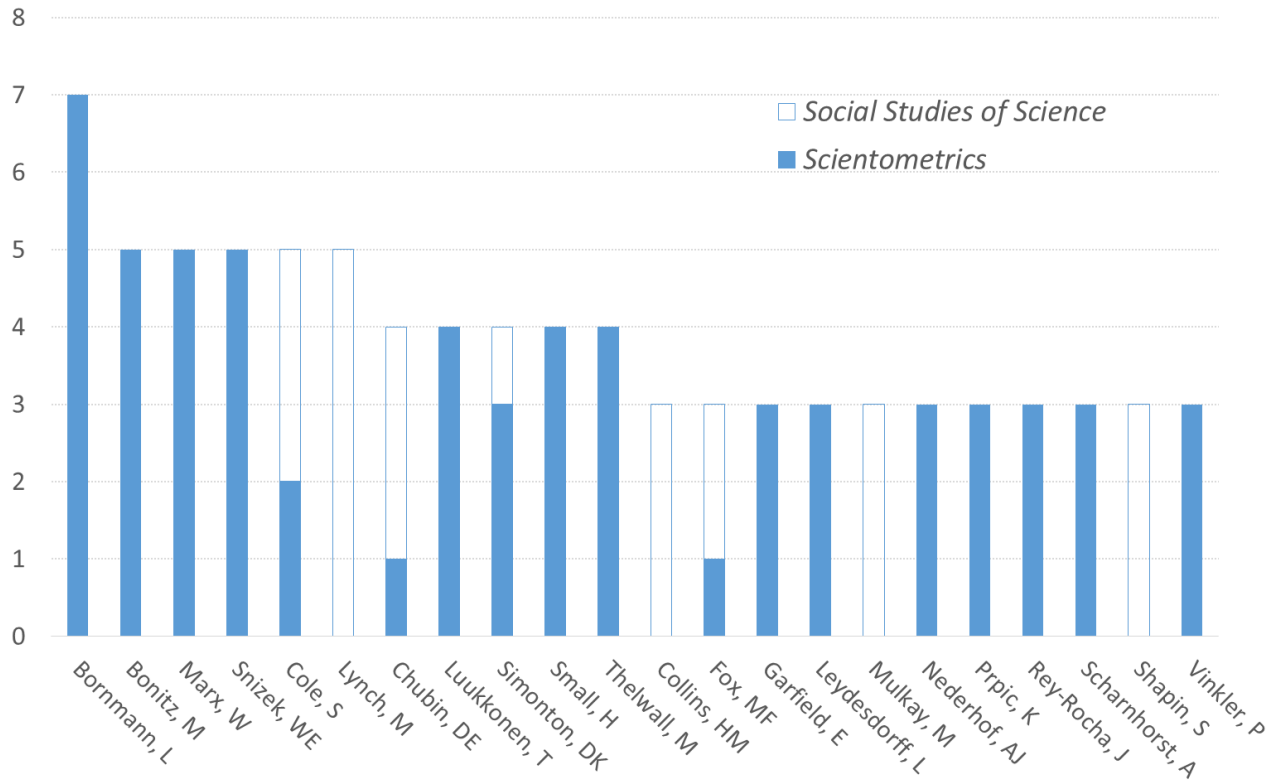

**Figure 7**: Eighty-five publications of twenty-two authors citing Merton as a first author (more than twice) in *Scientometrics* and *Social Studies of Science.*

Figure 7 shows the authors who cited Merton more than twice in *Scientometrics* or *Social Studies of Science.* Whereas most of these authors published exclusively in one of the two journals, Stephen Cole, one of Merton's students, has been prolific in both domains. Mary Frank Fox and Daryl Chubin also crossed the boundary. Other authors (e.g., Small, Garfield, and Leydesdorff) published in both journals, citing Merton when contributing to *Scientometrics*, but not when writing for *Social Studies of Science*. Others wrote exclusively for one of the two journals.

In summary: the name "Merton" as a concept symbol has obtained a different meaning in these two contexts of journals. On the sociological side, Merton has become a background figure who is cited incidentally. Garfield (1955) coined the term "obliteration by incorporation" (Cozzens, 1989): one no longer has to cite Merton explicitly and citation gradually decreases. McCain's (2015) noted that "obliteration by incorporation" is discipline dependent. In *Scientometrics*, however, the call for more theoretical work in addition to the methodological character of the journal has made referencing to Merton convenient.

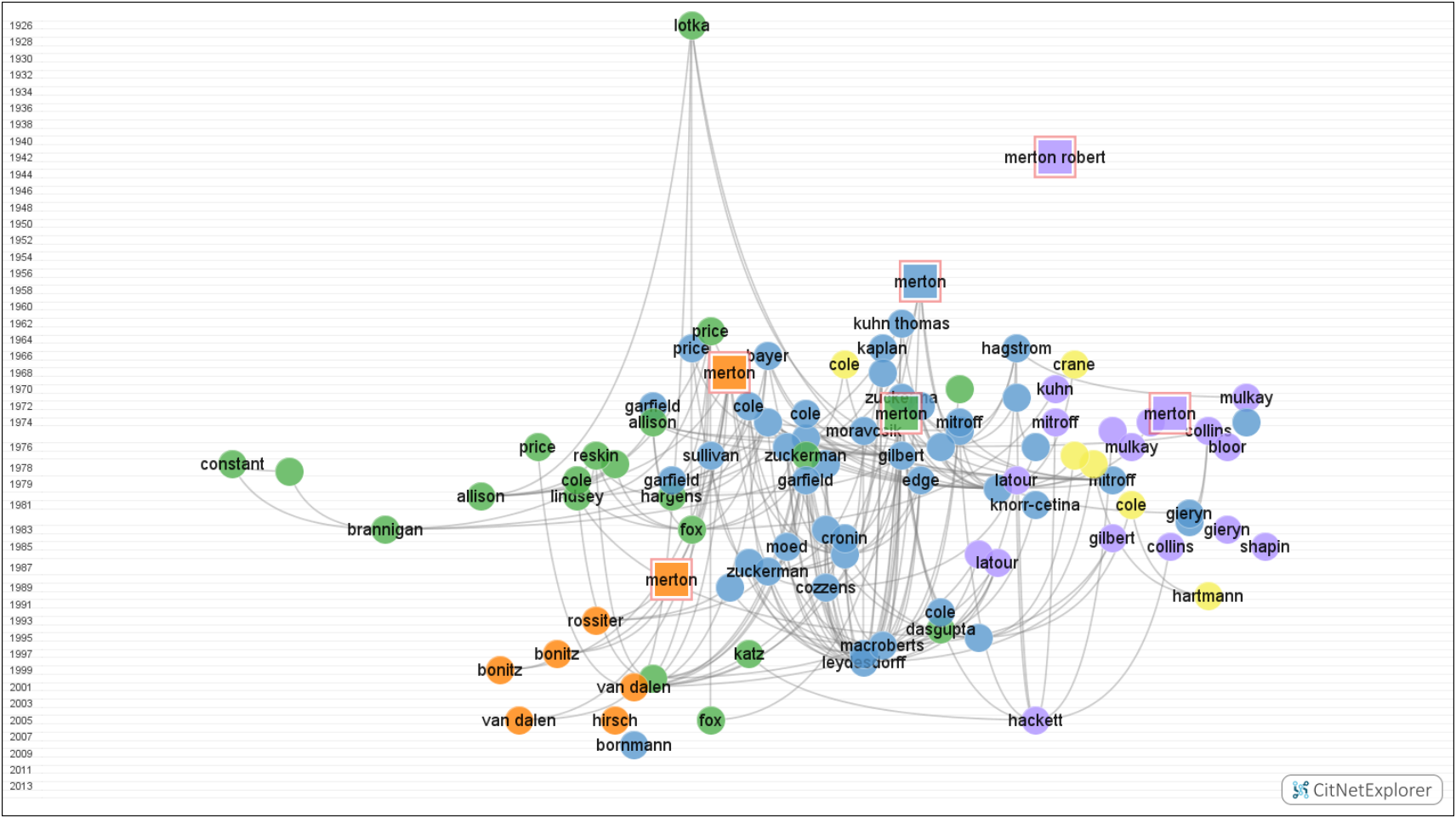

**Figure 8**: Citation network of 100 (of the 391) documents citing Merton in *Social Studies of Science* and *Scientometrics*. CitNetExplorerer used for the visualization.

Figure 8 shows, among other things, the influence of Merton's papers across the domains of quantitative and qualitative STS (Milojević *et al.*, 2014). Merton's (1973a) "Sociology-of-Science" book, for example, is positioned (on the right side) among the articles of qualitatively oriented sociologists like Harry Collins, Mike Mulkay, and David Bloor. Merton's (1988) paper about "Cumulative advantage and the symbolism of intellectual property" (also known as "Matthew II") is positioned on the other side among scientometricians, whereas several other references are to older work used in both traditions (e.g., Merton, 1957 and 1968).

In summary, Figure 8 visualizes the interface between the two branches of STS in terms of co-citation patterns. The integration by "Merton" as a concept symbol bridges the historically deep divide in terms of journals and institutions (e.g., Van den Besselaar, 2001). In other words, the figure illustrates the point that the different classifications of the two journals—*Scientometrics* as "library and information science"[15] and *Social Studies of Science* as "history and philosophy of science"—may cut through important elements of the intellectual organization of a field.

## 5. Multi-RPYS

We can illustrate our thesis of the two different functions of citation, at a research front or as longer-term codification, by using the Multi-Referenced Publication Years Spectroscopy (Multi-RPYS) recently introduced by Comins & Leydesdorff (2016). Multi-RPYS is an extension of RPYS, firstly introduced by Marx & Bornmann (2013) and Marx *et al.* (2014). In conventional RPYS one plots the number of references against the time axis. Figure 9 shows the result for the case of the 3,777 articles published in *Scientometrics* between 1978 and 2015.[16] The graph shows the numbers of yearly citations normalized as deviations from the five-year moving median. In this figure, for example, a first peak is indicated for 1926, indicating Lotka's (1926) Law as a citation classic in this field.[17]

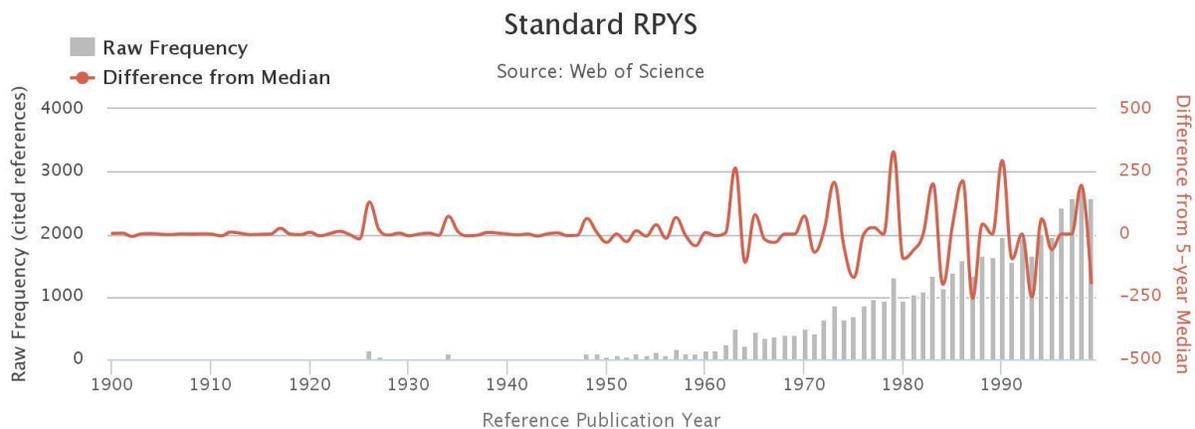

**Figure 9**: RPYS of 3,777 articles published in *Scientometrics*, downloaded on Jan. 2, 2016; curve generated using the interface at http://comins.leydesdorff.net .

---

[15] WoS classifies *Scientometrics* additionally as "Computer Science, Interdisciplinary Applications."
[16] Downloaded on Jan. 2, 2016.
[17] This curve was further analyzed in considerable detail by Leydesdorff *et al.* (2014).

CRExplorer < at http://www.crexplorer.net > enables the user to refine Figure 9 by disambiguating the cited references (Thor *et al.*, 2016). Elaborating on Comins & Husey (2015), Comins & Leydesdorff (2016) developed Multi-RPYS. Multi-RPYS maps RPYS for a series of years as a heat map. Figure 10 provides the Multi-RPYS for the same set of 3,777 articles from *Scientometrics.*

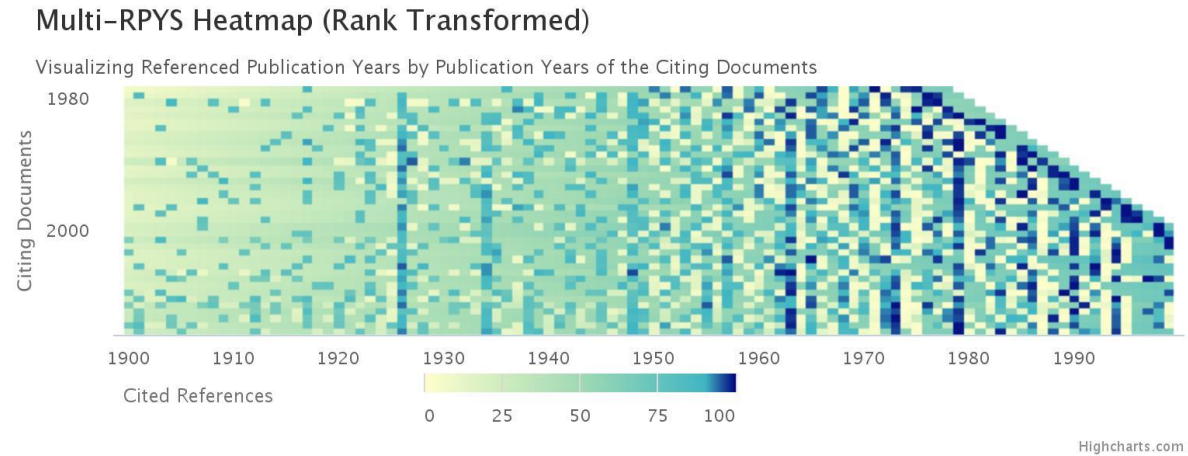

**Figure 10**: 3,777 articles published in *Scientometrics*, downloaded on Jan. 2, 2016. Source: http://comins.leydesdorff.net

**Table 2:** Ten most-cited publications in *Scientometrics* (before and after machine disambiguation using CRExplorer); Jan. 2, 2016

| Without disambiguation | | After disambiguation (including volume and page numbers)[18] | |
|---|---|---|---|
| hirsch je, 2005, p natl acad sci usa, v102, p16569 | 306 | hirsch je, 2005, p natl acad sci usa, v102, p16569 | 308 |
| de solla price d. j., 1963, little sci big sci | 171 | de solla price d. j., 1963, little sci big sci | 175 |
| lotka a. j., 1926, j washington acad sc, v16, p317 | 135 | garfield e., 1979, citation indexing | 154 |
| small h, 1973, j am soc inform sci, v24, p265 | 130 | lotka a. j., 1926, j washington acad sc, v16, p317 | 136 |
| katz js, 1997, res policy, v26, p1 | 125 | small h, 1973, j am soc inform sci, v24, p265 | 130 |
| garfield e, 1972, science, v178, p471 | 113 | katz js, 1997, res policy, v26, p1 | 128 |
| egghe l, 2006, scientometrics, v69, p131 | 108 | garfield e, 1972, science, v178, p471 | 113 |
| price djd, 1965, science, v149, p510 | 108 | price djd, 1965, science, v149, p510 | 109 |
| schubert a, 1986, scientometrics, v9, p281 | 106 | egghe l, 2006, scientometrics, v69, p131 | 108 |
| merton rk, 1968, science, v159, p56 | 105 | schubert a, 1986, scientometrics, v9, p281 | 106 |

Figure 10 shows the same bar in 1926, and similarly bars in 1963, 1973, 1979, etc. (Table 2). One can also see that citation of 1963 as referenced publication year became less intensive during the 1990s than in more recent years. Referencing to Price (1965), however, seems to have been obliterated by incorporation. On the top-right side of the figure, the progression of citing years generates an oblique cut-off. Two years behind this edge the dark blue blocks represent citation currency at the research front.

---

[18] CRExplorer allows for further disambiguation manually (Thor *et al.*, 2016).

Using *Gene* as a biomedical journal with a focus on the research front (Baumgartner & Leydesdorff, 2014, pp. 802f.), Figure 11 shows the predominance of the research front over longer-term citation in this case. However, the bars indicating longer-term citation are far from absent. The top-10 most-highly-cited papers (Table 3) are all more than ten years old.

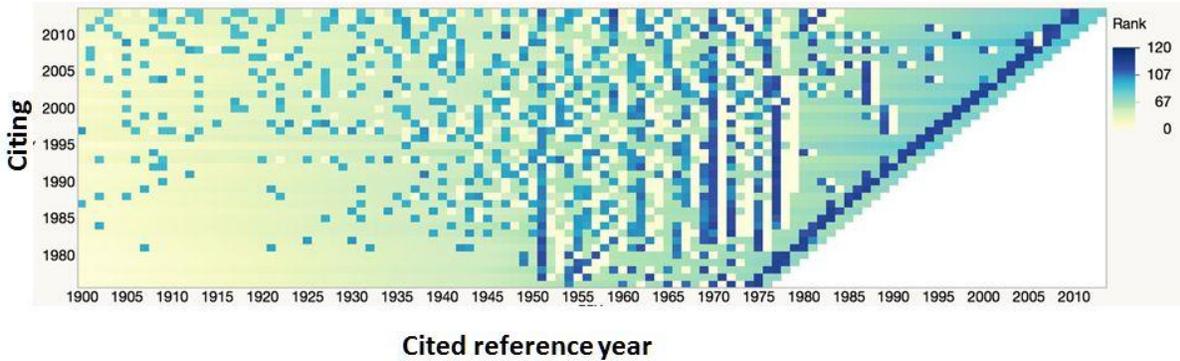

**Figure 11**: 15,383 articles published in *Gene* 1996-2015, downloaded on Feb. 14, 2016; visualized using the statistical software JMP.

**Table 3**: Ten most-cited publications in *Gene* (without disambiguation using CRExplorer); Feb. 14, 2016

| CITED REFERENCES | RPY | N_CR |
|---|---|---|
| sanger f, 1977, p natl acad sci usa, v74, p5463 | 1977 | 1718 |
| sambrook j., 1989, mol cloning lab manu | 1989 | 1272 |
| laemmli uk, 1970, nature, v227, p680 | 1970 | 818 |
| maniatis t., 1982, mol cloning | 1982 | 761 |
| thompson jd, 1994, nucleic acids res, v22, p4673 | 1994 | 616 |
| maniatis t, 1982, mol cloning laborato | 1982 | 609 |
| southern em, 1975, j mol biol, v98, p503 | 1975 | 609 |
| yanischperron c, 1985, gene, v33, p103 | 1985 | 544 |
| altschul sf, 1990, j mol biol, v215, p403 | 1990 | 504 |
| maxam a m, 1980, methods enzymol, v65, p499 | 1980 | 503 |

Figure 12 completes our argument by showing the results of Multi-RPYS for *Soziale Welt.* The research front is not present in all years, and also otherwise citation is not well organized in this journal. With a single exception, the top-10 most highly cited references are in German (Table 4).

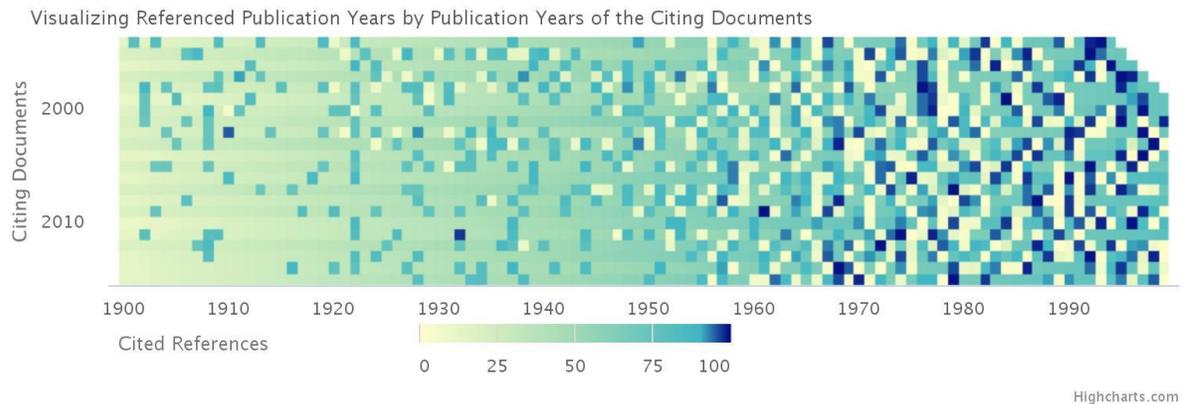

**Figure 12**: 524 documents published in *Soziale Welt* since its first edition in 1949; downloaded on Feb. 14, 2016. Source: http://comins.leydesdorff.net.

**Table 4**: Ten most-cited publications in *Soziale Welt* (automatic disambiguation using CRExplorer); Feb. 14, 2016

| CITED REFERENCES | RPY | N_CR |
|---|---|---|
| beck u., 1986, risikogesellschaft w | 1986 | 65 |
| luhmann niklas, 1984, soziale systeme | 1984 | 32 |
| luhmann niklas, 1997, gesellschaft gesells | 1997 | 30 |
| luhmann n, 2000, org entscheidung | 2000 | 23 |
| beck ulrich, 1993, erfindung politische | 1993 | 22 |
| schulze g., 1992, erlebnisgesellschaft | 1992 | 22 |
| meyer jw, 1977, am j sociol, v83, p340, doi 10.1086/226550 | 1977 | 21 |
| giddens anthony, 1995, konsequenzen moderne | 1995 | 19 |
| bourdieu p., 1982, feinen unterschiede | 1982 | 18 |
| beck ulrich, 2004, kosmopolitische euro | 2004 | 18 |

## 6. Summary and conclusions

We argue that the measurement of "quality" in terms of citations can further be qualified: one can, and probably should, distinguish between short-term citation currency at the research front and longer-term processes of the incorporation and codification of knowledge claims into bodies of knowledge. The latter can be expected to operate selectively, whereas the former provide variation. Citation impact studies focus on short-term citation, and therefore tend to measure not epistemic quality but involvement in current discourses and sustained visibility (Moed *et al*., 1985).

Major sources of data and a majority of the indicators used for the evaluation of science and scientists, are biased towards short-term impact. The use of Journal Impact Factors, for example, can be expected to lead to a selection bias that is skewing the results of evaluations in favor of short-term impact. The assumption of the existence of a research front underlying JIF and many other policy-relevant indicators (Price, 1970) is backgrounded in evaluation studies. However, in this study we have shown that even in the fields with a research front (exemplified in terms of short-term citations), there is significant presence of long-term citations. This calls for more studies (both theoretical and evaluative) examining the relationship between short- and long-term impact. We have shown that patterns emerging from multi-RPYS visualizations enable distinguishing between short-term and long-term impact (e.g., trailing edges and vertical bands, respectively).

The two processes of citation currency and citation as codification can be distinguished analytically, but they are coupled by feedback and feed-forward relations which evolve dynamically. At each moment of time, selection is structural; but the structures are also evolving, albeit at a slower pace. The dynamics of science and technology are continuously updated by variation at the research fronts. However, the relative weights of the processes of variation, selection, and retention can be expected to vary among the disciplines. There is no "one size fits all" formula. Research styles, disciplinary backgrounds, and methodological styles can be expected to vary within institutional units (e.g., departments, journals, etc.). A later concept symbol does not have to be prominently cited at the research front during the first few years (Ponomarev *et al*., 2014a and 2014b; cf. Baumgartner & Leydesdorff, 2014; Ke *et al*., 2015). Using a sample of 40 Spanish researchers, however, Costas *et al*. (2011) found that such occurrences (coined "the Mendel syndrome" by these authors) are rare. Baumgartner & Leydesdorff (2014) estimated that between five and ten percent of the citation patterns are atypical.

The transformation of the citing distribution into the cited one first generated an illusion of comparability (Wouters, 1999), but the normalization is based on assumptions about similarities in citing behavior without sociological reflection (Hicks *et al*., 2015). The citation distributions ("cited") thus generated are made the subject of study in a political economy of research evaluations (Dahler-Larsen, 2014) with the argument that one follows "best practices." We call this a political economy because the evaluations are initiated by and may have consequences for funding decisions; the production of indicators itself has become a quasi-industry. As we have shown in case of a German-language sociology journal, *Soziale Welt*, studying both citing and cited environments of the entity we focus on (individuals, groups, or journals) will be not only more informative in the studies of science via citations, but necessary in deciding reference sets for evaluative purposes. This becomes especially important in evaluation exercises of non-US (and non-English language) based research.

Are there alternatives? First, the processes of codification of knowledge via long-term citations can be studied empirically by expanding the focus from references (as the currently only way of measuring impact) to the full text of the published research as well. In the full texts, one can study the processes of obliteration by incorporation (e.g., McCain, 2015) and the different functions of referencing in arguments (e.g., Amsterdamska & Leydesdorff, 1989; Leydesdorff & Hellsten, 2005). The increased availability of full text with advances in textual analyses (e.g., Cabanac, 2014; Milojević, 2015) and citation-in-context studies (e.g., Small, 1982; 2011) are promising venues for further research.

Secondly, the dynamics of structure/agency contingencies is relevant to citation analysis (Giddens, 1989; Leydesdorff, 1995b). Citing can be considered as an action in which the author integrates cognitive, rethorical, and social aspects or, in other words, reproduces an epistemic, textual, and social dynamics. The structures of the sciences to which one contributes by reproducing them in instantiations (Fujigaki, 1998) are ideational and therefore latent; they are only partially reflected by individual scholars in specific texts. The texts make the different dynamics amenable to measurement (Callon *et al*., 1983; Leydesdorff, 1995a; Milojević, 2015).

From a structuralist perspective, the references in the texts can be modeled as variables contributing to the explanation of the dynamics of science and technology. Citations are then not reified as facts naturalistically found in and retrieved from databases. Whereas the indicators seem to be in need of an explanation (e.g., in a so-called "theory of citation"), considering this data as proxies of variables in a model turns the tables: not the citations need to be explained, but the operationalization of the variables in terms of citations has to be specified. From this perspective, issues such as normalization become part of the elaboration of a measurement theory (which is always needed). A scientometric research program can thus be formulated in relation to the history and philosophy of science (Comins & Leydesdorff, in preparation; cf. Leydesdorff, 1995a). However, the research questions about "quality" on this research agenda differ in important respects from those raised in evaluation studies about short-term impact.

**Acknowledgement**

We are grateful to Thomson Reuters for providing us with JCR data.